\documentclass[%
 reprint,
superscriptaddress,
 amsmath,amssymb,
 aps,
pra,
]{revtex4-1}

\usepackage{siunitx}
\usepackage{xcolor}
\usepackage[super]{nth}
\usepackage{graphicx}% Include figure files
\usepackage{dcolumn}% Align table columns on decimal point
\usepackage{bm}% bold math
\usepackage{amsmath,bm,braket}
\graphicspath{ {./images/} }
\usepackage{hyperref}
\hypersetup{colorlinks=true,
	linkcolor=red,
	citecolor=blue,
	urlcolor=orange}
\begin{document}

\preprint{APS/123-QED}

\title{Extending the Variational Quantum Eigensolver to Finite Temperatures}% Force line breaks with \\

\author{Johannes Selisko}
\email{johannes.selisko@de.bosch.com}
 \affiliation{Robert Bosch GmbH, Robert-Bosch-Campus 1, 71272 Renningen, Germany}
 \affiliation{ICAMS, Ruhr-Universit\"at Bochum, Universit\"atstr. 150, 44801 Bochum, Germany}%Lines break automatically or can be forced with \\
 
\author{Maximilian Amsler}%
\email{maximilian.amsler@de.bosch.com}
 \affiliation{Robert Bosch GmbH, Robert-Bosch-Campus 1, 71272 Renningen, Germany}

\author{Thomas Hammerschmidt}%
\affiliation{ICAMS, Ruhr-Universit\"at Bochum, Universit\"atstr. 150, 44801 Bochum, Germany}

\author{Ralf Drautz}%
\affiliation{ICAMS, Ruhr-Universit\"at Bochum, Universit\"atstr. 150, 44801 Bochum, Germany}

\author{Thomas Eckl}%
\email{thomas.eckl@de.bosch.com}
\affiliation{Robert Bosch GmbH, Robert-Bosch-Campus 1, 71272 Renningen, Germany}

\date{\today}% It is always \today, today,

\begin{abstract}
We present a variational quantum thermalizer (VQT), called quantum-VQT (qVQT), which extends the variational quantum eigensolver (VQE) to finite temperatures. The qVQT makes use of an intermediate measurement between two variational circuits to encode a density matrix on a quantum device. A classical optimization provides the thermal state and, simultaneously, all associated excited states of a quantum mechanical system. We demonstrate the capabilities of the qVQT for two different spin systems. First, we analyze the performance of qVQT as a function of the circuit depth and the temperature for a 1-dimensional Heisenberg chain. Second, we use the excited states to map the complete, temperature dependent phase diagram of a 2-dimensional J1-J2 Heisenberg model. The numerical experiments demonstrate the efficiency of our approach, which can be readily applied to study various quantum many-body systems at finite temperatures on currently available NISQ devices.
\end{abstract}

\maketitle

%\tableofcontents

\section{Introduction}
Recent advances in quantum computing have been driving intense research in the development of quantum algorithms that offer significant advantage over their classical counterparts. In particular, quantum algorithms are used for studying interacting many-electron systems that fundamentally govern the properties of materials and molecules~\cite{collaborators*_hartree-fock_2020, huggins_unbiasing_2022}. In quantum chemistry and materials modeling, a quantum advantage~\cite{lloyd_universal_1996} could be achieved either by offering a significant, potentially exponential acceleration of conventional methods to (approximately) solve the electronic Schr\"odinger equation, or by improving accuracy by incorporating a better description of the many-body effects of strongly correlated electronic systems~\cite{bauer_hybrid_2016}. However, universal fault-tolerant quantum hardware~\cite{shor_fault-tolerant_1996} is required to harness the full potential of quantum computing, which is expected to be deployed only within the next decade. Currently available experimental devices, so-called Noisy Intermediate-Scale Quantum computers (NISQ), are limited by their inherent circuit noise and their decoherence time, posing strong constraints with respect to the number of qubits, the circuit depth, and the number of gate operations which can be executed within quantum algorithms~\cite{preskill_quantum_2018}.

Due to these constraints, the execution of algorithms like quantum phase estimation or quantum Fourier transform are impractical on NISQs, and most methods that produce quantum circuits executable on available hardware are centered around hybrid quantum-classical algorithms. For example, the variational quantum eigensolver (VQE)~\cite{peruzzo_variational_2014, tilly_variational_2021, cerezo_variational_2021} uses a quantum computer to store a parametrized wave function and measure its energy, while a classical, external minimization of the energy through variational parameters provides an approximation to the ground-state of the system. Although the classical optimization in a VQE is challenging due to the presence of local minima and barren plateaus~\cite{mcclean_barren_2018,wang_noise-induced_2021,cerezo_variational_2021}, its utility has already been experimentally demonstrated for small molecular systems~\cite{collaborators*_hartree-fock_2020}.

%In this paper we demonstrate the principle of the extension of VQE to thermal states. We call our algorithm full variational quantum thermalizer (FVQT) as it embeds both the entropy and energy calculation on the quantum device.\\
%The next subsection explains the principle of thermal states and current quantum algorithms to compute the Gibbs state of a system.
%Then we present the algorithm and analyze its cost in terms of measurement precision, memory requirements and complexity. Further on we show two applications of FVQT, the first is a one-dimensional Heisenberg chain with external fields, where we have a look at the required circuit depth to correctly describe the system and the temperature dependence of the algorithm. In the second example, the two dimensional J1-J2 Heisenberg model we calculate the phase diagram from the excited states obtained by FVQT.

%\subsection{Calculating thermal properties on quantum computers}

In addition to the ground states, many applications require the assessment of the thermal state at finite temperatures (Gibbs state) or the properties of excited states in a system. The Gibbs state minimizes the Helmholtz free energy $F = E-TS$ at an inverse temperature $\beta=1/T$ (in units of $k_B^{-1}$) with the energy $E$ and the entropy $S$. This state is mixed and can be formulated in terms of the eigenstates $\ket{\varphi_i}$ and eigenenergies $\epsilon_i$ of the Hamiltonian:
\begin{align}
\hat{\rho}_\text{Gibbs} & = \sum_{i=0} p_i\ket{\varphi_i}\bra{\varphi_i},
\label{eq:definition_Gibbs_state}
\end{align}
with
\begin{align}
p_i = \frac{e^{-\beta \epsilon_i}}{Z},\qquad Z &=  \sum_{i} e^{-\beta \epsilon_i}.
\label{eq:partition_function}
\end{align}

To date, two classes of algorithms have been developed to compute the Gibbs state. The first class is based on computing each eigenstate separately and subsequently mixing them according to their probabilities $p_i$ in equation~\eqref{eq:definition_Gibbs_state}. Such algorithms usually start out by computing the ground-state using a VQE, followed by successively computing the excited (eigen-) states and projecting them out, or penalizing the already computed eigenstates~\cite{kuroiwa_penalty_2020,higgott_variational_2018}.

The second class prepares the Gibbs state itself on the quantum device, which involves the explicit treatment of the entropy $S$. The thermofield-double-states method, for example, doubles the system and collapses it in order to introduce entropy into the system~\cite{wu_variational_2018,wang_variational_2020,sagastizabal_variational_2021}. However, the measurement of the entropy is far from trivial: a quantum computer can measure the expectation value of an operator $\hat{O}$ on a state described by a density matrix $\hat{\rho}$ by taking the trace $\text{Tr}(\hat{O}\hat{\rho})$, but the expression for the entropy $\hat{\rho}\ln(\hat{\rho})$ is not linear in $\hat{\rho}$, rendering the corresponding measurement more demanding. On the other hand, using imaginary time evolution to obtain the Gibbs state requires complex quantum circuits which are challenging to implement on NISQ devices~\cite{motta_determining_2020,nishi_implementation_2021}. Verdon \textit{et al.}~\cite{verdon_quantum_2019} recently introduced a combination of a machine-learning algorithm with a variational quantum circuit, called hybrid variational quantum thermalizer (hVQT), which generalizes the VQE towards finite temperatures and involves a neural network that learns the entropic probability distribution, while a quantum circuit prepares the eigenstates of the Hamiltonian (see also Ref.~\onlinecite{guo_thermal_2021}). However, this approach leads to an intimate interaction of classical and quantum computation, which typically results in longer runtimes, and has thus motivated the development of alternate algorithms~\cite{foldager_noise-assisted_2022}.

To alleviate above issues of the hVQT and enable the algorithm to maximally benefit from a possible advantage of quantum machine-learning~\cite{Perdomo_Ortiz_2018,Alcazar_2020} we develop an algorithm which transfers the generation of the entropic probability distribution directly onto the quantum computer, thereby allowing to fully assess the Gibbs state in the quantum device. This approach, which we call quantum-VQT (qVQT), offers significant advantages over the hVQT: it minimizes the communication between classical and quantum computer and is able to achieve accurate results using significantly fewer measurements by evaluating them according to the probability distribution. In the remainder of this manuscript we present in detail the qVQT algorithm and demonstrate its performance by applying it to solve a 1-dimensional Heisenberg chain, and by computing the complete, temperature dependent phase diagram of a 2-dimensional J1-J2 Heisenberg model.

\section{Method}

\begin{figure}
	\includegraphics[scale=0.3]{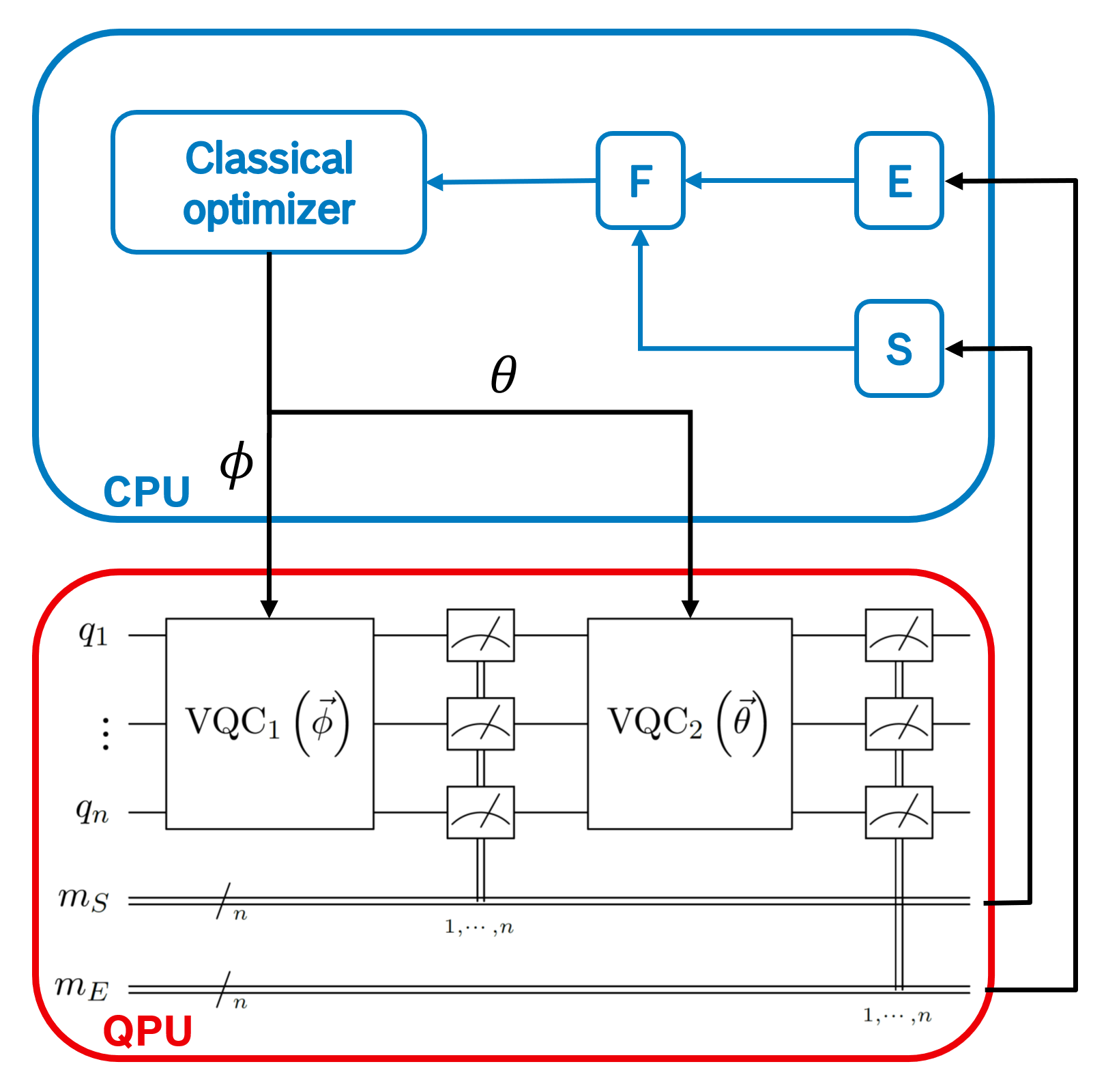}% Here is how to import EPS art
	\caption{\label{fig:FVQT} A schematic illustration of the qVQT algorithm, showing in blue the part which is executed on the classical computer (CPU) and includes the calculation of energy $E$ and entropy $S$ to form the free energy $F$ as well as a classical optimization routine. The red part shows the quantum circuit which is evaluated using the quantum computer (QPU). It consists of two variational circuits with an intermediate measurement for obtaining the entropy and a final measurement to obtain the energy.}
\end{figure}

\subsection{Principles of the qVQT}

The flowchart and the relevant components of the qVQT algorithm are shown in Fig.~\ref{fig:FVQT}. The fundamental idea of the qVQT is to use two separate variational quantum circuits (VQC) and an intermediate measurement to obtain a mixed state on a quantum computer (red block ``QPU'' in Fig.~\ref{fig:FVQT}). A classical optimization determines the parameters for which this mixed state represents the Gibbs state (blue block ``CPU'' in Fig.~\ref{fig:FVQT}). Different flavors of VQC have been proposed in the literature, e.g., hardware efficient VQC, particle number conserving VQC, variational Hamilton Ansatz (VHA), etc., which can also be used in qVQT~\cite{tilly_variational_2021,fedorov_vqe_2022,gard_efficient_2020}. We denote the parameters of the first VQC ($\text{VQC}_1$) with $\vec{\displaystyle \phi}$, while the parameters of the second VQC ($\text{VQC}_2$) are referred to as $\vec{\displaystyle\theta}$.

The first variational circuit $\text{VQC}_1\big(\vec{\displaystyle \phi}\big)$ and the intermediate measurement generate a classical distribution (see ``QPU'' in Fig.~\ref{fig:FVQT}). Specifically, the superposition of the basis states $\ket{b_i}$ produced by $\text{VQC}_1\big(\vec{\displaystyle \phi}\big)$ collapses to a probability distribution. $\hat{\rho}_{\text{VQC}_1}$ in equation~\eqref{eq:states_VQC1_midterm_a} represents the density matrix after the first variational circuit $\text{VQC}_1\big(\vec{\displaystyle\phi}\big)$, while $\hat{\rho}_{mm}$ in equation~\eqref{eq:states_VQC1_midterm} is the density matrix after the intermediate measurement:
\begin{align}
\hat{\rho}_{\text{VQC}_1} &= \left(\sum_{i} a_i\big(\vec{\displaystyle\phi}\big) \ket{b_i}\right)\left(\sum_{i} a_i^*\big(\vec{\displaystyle\phi}\big) \bra{b_i}\right) 
\label{eq:states_VQC1_midterm_a}
\\
\hat{\rho}_{mm} &= \sum_{i} \left|a_i\big(\vec{\displaystyle\phi}\big)\right|^2 \ket{b_i}\bra{b_i}
\label{eq:states_VQC1_midterm}.
\end{align}

The second variational circuit $\text{VQC}_2\big(\vec{\displaystyle\theta}\big)$  maps the basis states $\ket{b_i}$ to a superposition of these basis states while preserving the orthogonality, and prepares the state
\begin{align}
\hat{\rho}_{\text{VQC}_2} &= \sum_{i} \left|a_i\big(\vec{\displaystyle\phi}\big)\right|^2 \ket{\psi_i\big(\vec{\displaystyle\theta}\big)}\bra{\psi_i\big(\vec{\displaystyle\theta}\big)}
\label{eq:states_VQC2}
\end{align}
which has the same form as equation~\eqref{eq:definition_Gibbs_state}. 

We obtain the free energy $F$ by minimizing its value over the parameter set $(\vec{\displaystyle\phi},\vec{\displaystyle\theta})$, a task which is performed classically using an arbitrary (local) optimizer (see ``CPU'' in Fig.~\ref{fig:FVQT}). The energy $E$ is obtained by measuring the expectation value of the Hamilton operator after the second variational circuit $\text{VQC}_2$, and the entropy $S$ is obtained by the intermediate measurement of $\text{VQC}_1$. From the probabilities $p_i$ of measuring the basis state $\ket{b_i}$ we obtain the entropy:
\begin{align}
S = \sum_{i} p_i\ln(p_i),	\label{eq:definition_entropy}
\end{align}

\subsection{Computational Cost}
To assess the resource cost and scaling of the qVQT we first discuss the error estimate as a function of the number of measurements in section~\ref{sec:precision}, then determine the required memory resources in section~\ref{sec:memory}, and finally analyze the complexity of the qVQT in section~\ref{sec:complexity}.

\subsubsection{Measurement Precision}\label{sec:precision}
Drawing $N$ samples from a random distribution with standard deviation $\sigma$ yields a standard error of $\Delta = \frac{\sigma}{\sqrt{N}}$.
%\paragraph{Measurement of Each Eigenstate}
Hence, an algorithm which measures all eigenvalue $\epsilon_i$ of a Hamiltonian yields a standard error of $\Delta \epsilon_i = \sigma_i/\sqrt{N/2^n}$ for each of these measurements, where $\sigma_i$ is the standard deviation of the measurement of the eigenvalue $\epsilon_i$ and $n$ is the dimension of the system. The factor $2^n$ arises from splitting up the number of measurements among all $2^n$ eigenstates.
The energy of the Gibbs state
\begin{align}
	E = \frac{1}{Z}\sum_{i} e^{-\beta \epsilon_i}\epsilon_i = \sum_{i=0} p_i \epsilon_i
\end{align} 
will then have an approximate error of
\begin{align}
	\Delta E = \frac{1}{\sqrt{N}}\sqrt{2^n\sum_{i} \sigma_i^2p_i^2\left(1-\beta \epsilon_i(1-p_i)\right)^2},\label{eq:error_a}
\end{align}
where the partition function and probabilities are computed from the energy measurements.

%\paragraph{Measurement of Thermal State}

Within the qVQT, we measure the expectation value of the thermal state directly with $N$ measurements. The precision depends on the measurement error $\Delta p_i$ of the probabilities and the measurement error $\Delta \epsilon_i$ of the eigenstates. The measurement error of $p_i$ can be calculated from the associated standard deviation, while the error of the energy eigenstate is given by the number of its measurements $p_iN$:
\begin{equation}
\begin{aligned}
	\Delta p_i &= \frac{\sqrt{p_i(1-p_i)}}{N}\\
	\Delta \epsilon_i &= \frac{\sigma_i}{\sqrt{p_iN}}.
\end{aligned}
\end{equation}
This leads to an error of the energy of the Gibbs state
\begin{align}
\Delta E = \frac{1}{\sqrt{N}}\sqrt{\sum_{i} \left[\sigma_i^2p_i+\frac{p_i(1-p_i)}{N}\epsilon_i^2\right]}\label{eq:error_b}
\end{align}
A detailed derivation of above relations can be found in appendix~\ref{sec:DervMeasPrec}.

%\paragraph{Comparison}

When comparing the two error estimates from equations~\eqref{eq:error_a} and~\eqref{eq:error_b}, we see that in both cases the leading order is $1/\sqrt{N}$. However, the pre-factor in~\eqref{eq:error_a} shows an additional energy term and an additional factor of $2^np_i$, which can increase the error dramatically. For example, in the limit of $\beta = 0$ we see that both cases yield the same pre-factor of the leading order. On the other hand, in the limit of $T=0$, the error in equation~\eqref{eq:error_b} decays exponentially compared to~\eqref{eq:error_a}, illustrating the advantage of the qVQT. 

\subsubsection{Memory Requirements}\label{sec:memory}
Computing the entropy requires the storage of all probabilities. Since the number of states grows exponentially with the system size the memory requirement $M$ grows exponentially as well up to the point where the number of measurements $N$ limits the number of states which are measured. This could be the case if there are excited states with probabilities comparable to $1/N$. Since the states and counts are stored in a dictionary, all states which do not occur in the $N$ measurements do not require any memory, all other states require memory $M(N)$ to store an integer smaller than $N$.

The qVQT can circumvent this issue if we allow the first variational circuit to split the system into $n/n_s$ independent subsystems of size $n_s$. In this case, we can determine the total entropy from the entropies of the individual subsystems, thereby allowing the algorithm to scale linearly with system size.

\begin{equation}
\begin{aligned}
M &= \frac{n}{n_s} 2^{n_s} \cdot M(N)
\end{aligned}
\end{equation}

\subsubsection{Complexity}\label{sec:complexity}
The goal of the qVQT is to approximate a density matrix of a mixed state, which is a $2^n\times 2^n$-dimensional complex and symmetric matrix with a trace of $1$ for a system of dimension $n$. Therefore, the qVQT has $2^{2n}-1$ degrees of freedom. An equivalent VQE only tries to find a pure state and has hence $2^{n+1}-2$ degrees of freedom. We assess whether the computational effort is comparable to a VQE with twice as many qubits or if it adds an exponential pre-factor to the cost of a VQE in Appendix \ref{sec:apx_scaling_chain}. For the rather small examples used in our numerical experiments we find that the necessary number of parameters, and therefore the cost of the optimization, highly depends on the variational circuits and we do not find evidence for an exponential scaling compared to VQE.

\section{Results and Discussions}
We demonstrate the utility of the qVQT by investigating two model systems: a 1-dimensional Heisenberg chain and a 2-dimensional J1-J2 Heisenberg model. For this purpose, we implement the qVQT algorithm using toolchains provided by qiskit~\cite{Qiskit} and its associated quantum simulator. For all numerical experiments we consider three performance metrics (similar to Ref.~\onlinecite{verdon_quantum_2019}). The first one is given by the difference between the numerically computed and the exact free energy, while the second one is given by the fidelity as
\begin{align}
	f(\hat{\rho}_1,\hat{\rho}_2) = \left(\text{Tr}\left(\sqrt{\sqrt{\hat{\rho}_1}\hat{\rho}_2\sqrt{\hat{\rho}_1}}\right)\right)^2.
\end{align}
To obtain a criteria which vanishes as the density matrix $\hat{\rho}$ approaches the Gibbs state $\hat{\rho}_\text{Gibbs}$ we use the metric $f_m=1-f(\hat{\rho},\hat{\rho}_\text{Gibbs})$.
The third metric we use is the trace distance:
\begin{align}
	Td(\hat{\rho},\hat{\rho}_\text{Gibbs}) = \frac{1}{2}\text{Tr}\left(\sqrt{(\hat{\rho}-\hat{\rho}_\text{Gibbs})^\dagger(\hat{\rho}-\hat{\rho}_\text{Gibbs})}\right)
\end{align}
All three metrics vanish in the limit of ideal performance.

\subsection{1D Heisenberg Chain with Transverse Fields}\label{sec:1dHB}
The first model system we investigate is the 1D Heisenberg chain with transverse fields and nearest neighbor hopping, given by the Hamiltonian:
\begin{align}
	H = \sum_{\braket{ij}} J\left[\sigma_i^x\sigma_j^x + \sigma_i^y\sigma_j^y + \sigma_i^z\sigma_j^z\right] + \sum_i \left[J_x\sigma_i^x+J_z\sigma_i^z\right],
\end{align}
where $\sigma_i^{\{x,y,z\}}$ denotes a Pauli $\{x,y,z\}$ operator on qubit $i$ and the first sum runs over all pairs $\braket{ij}$ of nearest neighbors.
Verdon \textit{et. al}~\cite{verdon_quantum_2019} analyzed this particular model for 4-qubits with $J=-1$, $J_z=0.2$ and $J_x=0.3$ using the hVQT. To allow a direct comparison we employ the same parameters and use the qVQT to calculate the thermal state at the inverse temperature of $\beta =1.3$, which Verdon \textit{et. al} pointed out to be the most challenging.

\subsubsection{Required Circuit Depth\label{sec:reqcirc}}

\begin{figure}[h!]
	\includegraphics[scale=0.6]{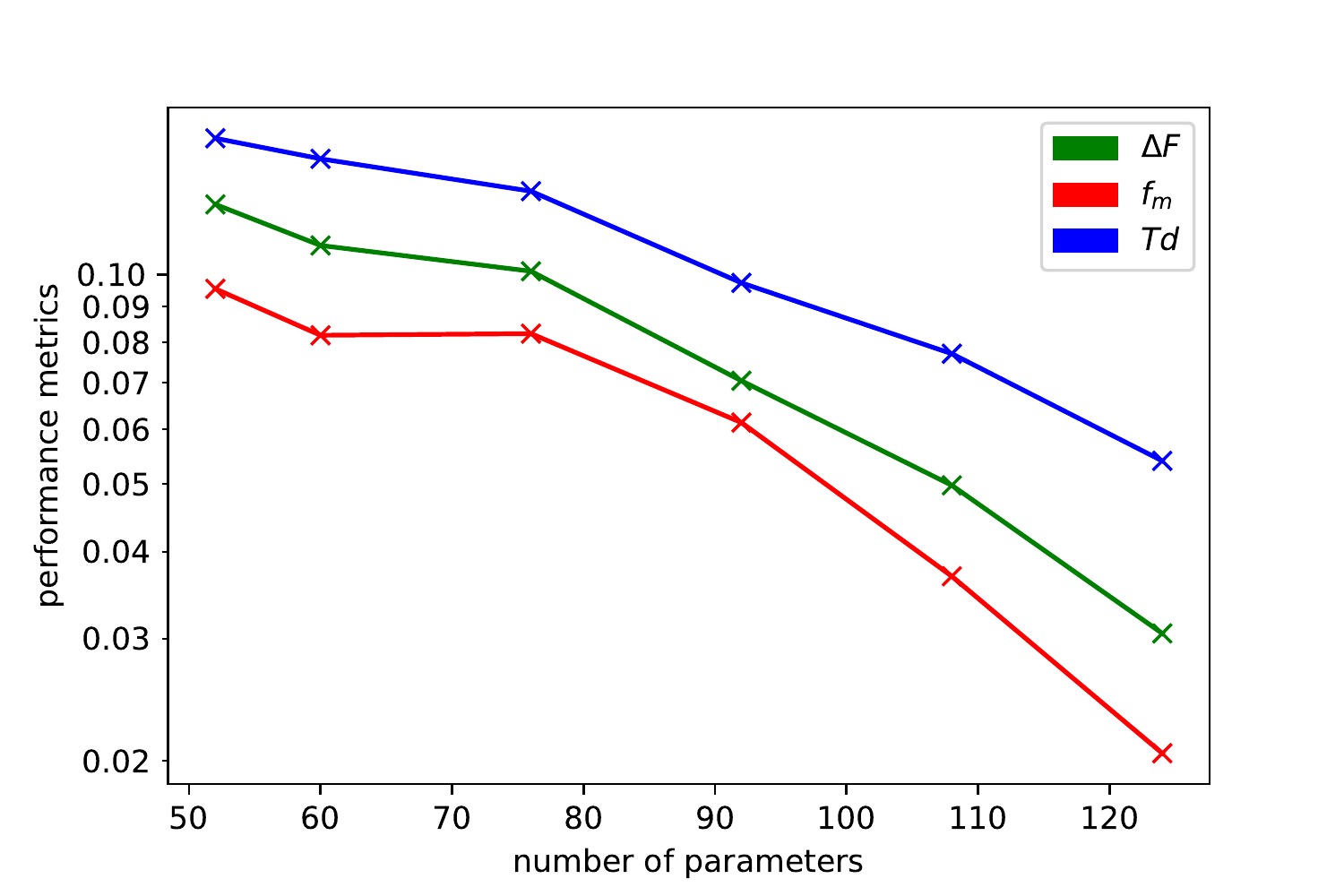}% Here is how to import EPS art
	\caption{\label{fig:4q_20perc_numvar} Results of a 4-qubit 1D Heisenberg chain with transverse fields at $\beta=1.3$, showing the top \nth{20} percentiles of the three performance metrics as a function of the total number of parameters. The difference between the computed and exact values of the free energy, the $f_m$-fidelity measure between the computed and exact density matrix, and the trace distance between the computed and exact density matrix are shown in green, red, and blue, respectively.}
\end{figure}

For a qVQT with a minimal entropy circuit, i.e., where the circuit $\text{VQC}_1$ only contains a Pauli $x$ rotation on each qubit, we already obtain accurate results with rapid convergence. Increasing the number of variational parameters for the energy circuit $\text{VQC}_2$ further improves accuracy. We perform a statistical analysis with different total number of variational parameters by conducting 100 runs starting from random initial parameters, and using the limited-memory Broyden-Fletcher-Goldfarb-Shanno bound optimizer (L\_BFGS\_B) \cite{Byrd_BFGS_1995,Zhu_BFGS_1997} as implemented in qiskit with a target gradient tolerance of \num{1e-3}. The three performance metrics of these statistical experiments are shown in Fig.~\ref{fig:4q_20perc_numvar}, illustrating that the qVQT yields an approximation to the density matrix that can be improved in accuracy by increasing the circuit depth and the associated computational cost. The corresponding scaling with respect to the required number of variational parameters is linear, as suggested by our numerical experiments (see appendix~\ref{sec:apx_scaling_chain}).

\subsubsection{Temperature Dependence}

\begin{figure}
	\includegraphics[scale=0.6]{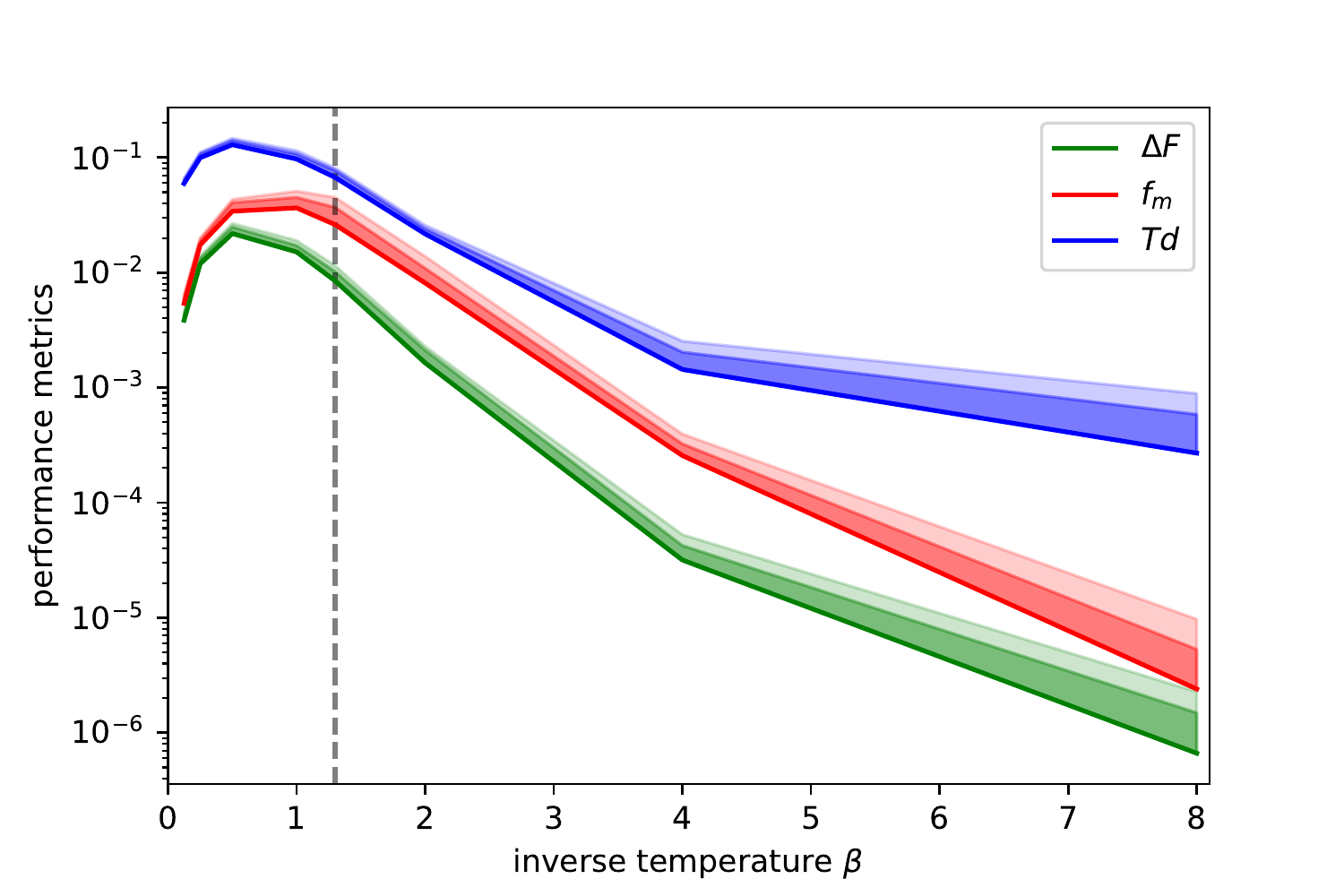}% Here is how to import EPS art
	\caption{\label{fig:crit_beta} The temperature dependence of a 4-qubit 1D Heisenberg chain with transverse fields, showing the three performance metrics as a function of the inverse temperature. The color coding corresponds to the one in Fig.~\ref{fig:4q_20perc_numvar}. The bottom line within each criteria correponds to the best result obtained from 100 runs starting from random parameters, using the L\_BFGS\_B optimizer with a gradient tolerance of \num{1e-3}. The dark shaded region denotes the range of the best 20-percentile of all runs, while the light shaded region denotes the interval up to the average over all runs. The dashed vertical line denotes the value of $\beta=1.3$ used in Sec.~\ref{sec:reqcirc}}
\end{figure}

In Fig.~\ref{fig:crit_beta} we show the dependence of the three performance metrics on the inverse temperature.  In the low-temperature limit ($\beta\gg 1$) the ground state is dominant and the accuracy improves as the algorithm does not need to calculate the excited states very precisely. In the high-temperature limit ($\beta\ll 1$), on the other hand, the splitting of all eigenstates becomes less important in comparison to the entropy. Clearly, temperatures around $\beta=1$ are the most challenging for the algorithm as both classical and quantum-mechanical correlation need to be correctly captured.

Since the qVQT algorithm is similar to a hVQT and mainly differs by the method to produce the classical probability distribution, the temperature dependence above is comparable to the results of Verdon \textit{et al.}~\cite{verdon_quantum_2019}. However, and most importantly, by switching from hVQT to qVQT a significantly smaller number of quantum circuits need to be executed on the quantum device due to the intermediate measurement, a key advantage of the qVQT.

\subsection{2D J1-J2 Heisenberg Model}

The second model system we investigate is the Heisenberg model with nearest and next-nearest neighbor interactions. Such systems have been extensively used to simulate and better understand the behavior of real magnetic materials that can be mapped to a spin model~\cite{noauthor_thermodynamic_2016}. The Hamiltonian of this model is given by:
\begin{equation}
	H = \sum_{\braket{i,j}} J_1\vec{S}_i\vec{S}_j+\sum_{\braket{\braket{i,j}}} J_2\vec{S}_i\vec{S}_j,
\end{equation}
where the first and second summations run over nearest and next-nearest neighbors, respectively. It is well known that this system develops three phases, depending on the relative interaction parameter $\alpha$, defined as $J_1 = \sin(\alpha)$ and $J_2=\cos(\alpha)$, which introduces a normalization of $J_1^2+J_2^2=1$.  When $J_2>0$ and the next nearest neighbor interactions are stronger than the nearest neighbor interactions, the spins form a \textit{stripe} configuration. When the nearest neighbor interactions are more important or $J_2<0$, the system turns ferromagnetic (FM) or antiferromagnetic (AFM) for $J_1<0$ and $J_1>0$, respectively.

To distinguish these three phases we construct two correlation functions that serve as order parameters:
\begin{align}
	c_0 &= \frac{1}{N_{nn}}\sum_{\braket{i,j}} \braket{\sigma_i^z\sigma_j^z}\nonumber\\
	c_1 &= \frac{1}{N_{nnn}}\sum_{\braket{\braket{i,j}}} \braket{\sigma_i^z\sigma_j^z},
\end{align}
where $c_0$ is the nearest neighbor correlation function averaged over all $N_{nn}$ pairs of nearest neighbors, and, analogously, $c_1$ is the next-nearest neighbor correlation function. Note that the Hamiltonian is symmetric for rotations of all spins in the same way, i.e., for every eigenstate also the rotated states are eigenstates of the Hamiltonian with the same energy. For the term used to calculate the correlation functions this translates to $\sigma_i^z\sigma_j^z = \vec{S}_i\vec{S}_j/3$. Because the eigenvalues of $\vec{S}_i\vec{S}_j$ range between $-3$ and $1$, the correlation functions can assume values in the interval $[-1,1/3]$.  A correlation function which is greater than zero means that the ferromagnetic contribution is dominating the antiferromagnetic one, while for a negative correlation function the antiferromagnetic contribution is dominant.

We perform qVQT experiments for a 4-qubit, two-dimensional J1-J2 Heisenberg lattice at a range of parameter angles $\alpha$ and an inverse temperature of $\beta=1$ to obtain the eigenstates of the Hamiltonian and, from those, the phase diagram. At each value of  $\alpha$ we perform 100 runs, starting with random initial parameters, again using the L\_BFGS\_B optimizer with a gradient convergence tolerance of \num{1e-3}. Our qVQT algorithm uses two hardware efficient variational circuits with depths of 2 and 7 for $\text{VQC}_1$ and $\text{VQC}_2$, respectively, which results in a total of 76 variational parameter.

\begin{figure}
	\includegraphics[width=0.90\columnwidth]{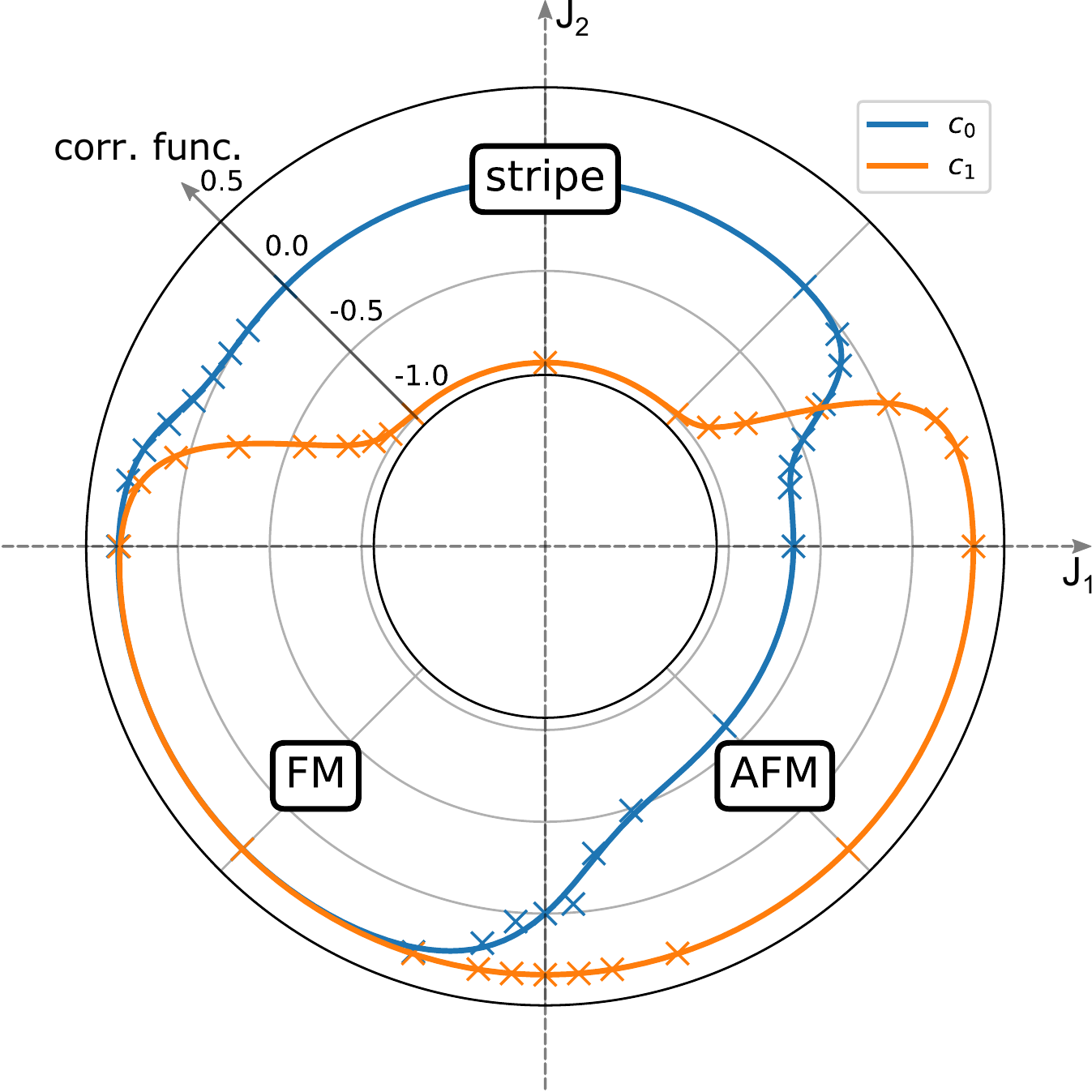}% Here is how to import EPS art
	\caption{\label{fig:corelpolar2} Nearest neighbor $c_0$ and next-nearest neighbor correlation function $c_1$ of a 4-qubit 2D J1-J2 Heisenberg model as a function of the parameter angle $\alpha$ at inverse temperature $\beta=1$. The lines are the results obtained by exact diagonalization, while the crosses denote the qVQT-results.}
\end{figure}

The exact correlation functions and the results obtained by the qVQT-experiments at $\beta = 1$ are shown in Fig.~\ref{fig:corelpolar2}. We clearly see the three phases together with the corresponding phase transitions. For almost all parameter angles the numerical experiments match the exact values very well, except for the region near the phase transition between the ferromagnetic and the antiferromagnetic phase at $J_1=0$ and $J_2=-1$. Remarkably, these increased errors seem to be of similar order on both sides of the transition and therefore do not influence the transition angle where the correlation function $c_0$ becomes zero, the point that we identify with the phase transition. Two possible explanations for this errors are that (a) either the approximation of the classical probability distribution breaks down in this regime, or (b) that the splitting of the energy eigenstates cannot be performed precisely when the ferromagnetic and antiferromagnetic states are in the same energy domain.

\begin{figure}[t!]
	\includegraphics[width=1.0\columnwidth]{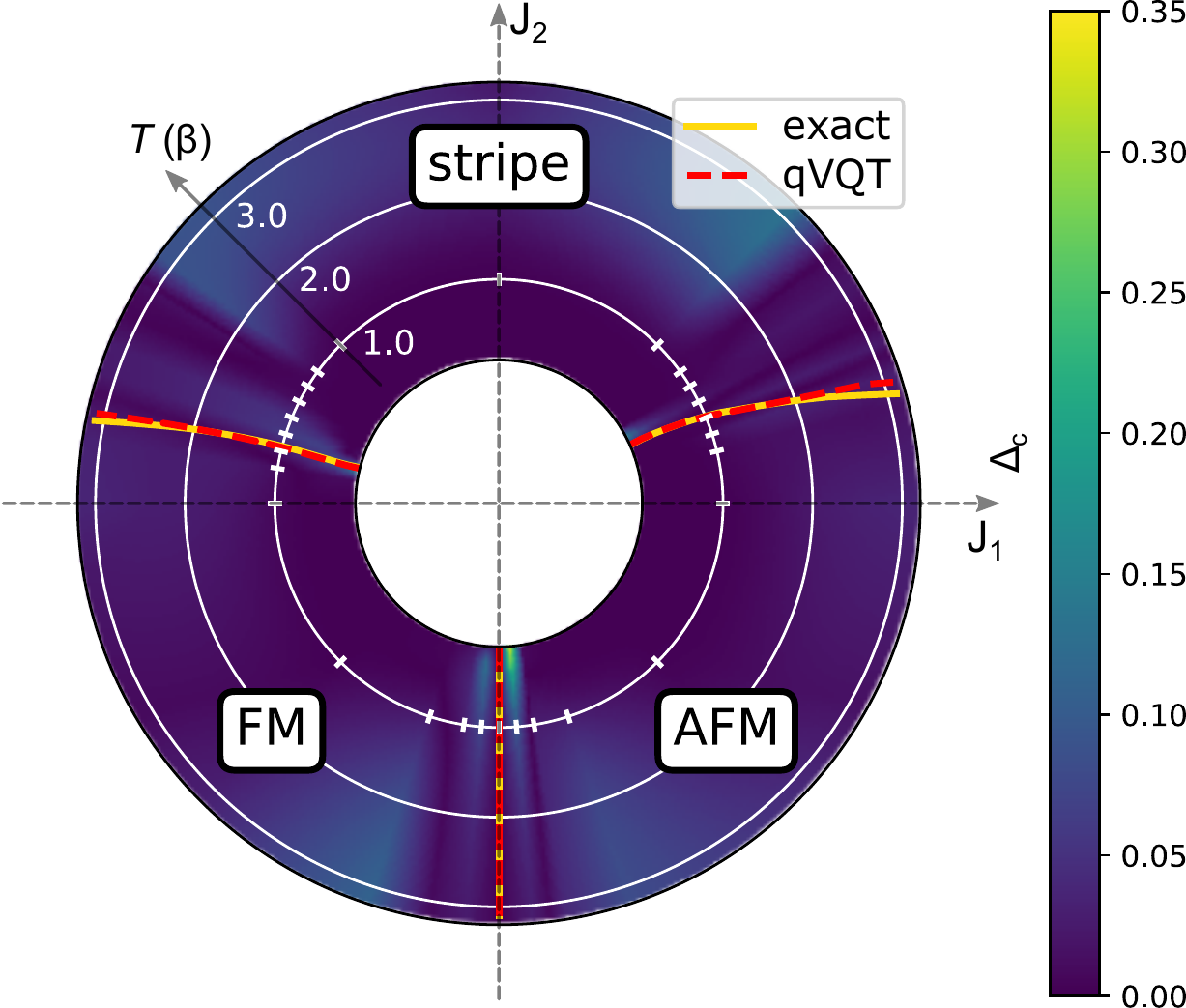}% Here is how to import EPS art
	\caption{\label{fig:phasediag_T_theoexp} The phase diagram of a 4-qubit 2D Heisenberg model, obtained from a qVQT experiment (dashed red) together with the exact ground truth (solid yellow). The phase transitions are given by the angles where the correlation functions vanish. The colormap indicates the combined deviation of the correlation functions qVQT from the exact results, $\Delta_c = |c_0^{theo}-c_0^{qVQT}|+|c_1^{theo}-c_1^{qVQT}|$, with linear interpolation between all data points. The explicit qVQT results are calculated on the white crosses at $T=1$ and classically extended to temperatures $T\neq1$.}
\end{figure}

Next, we compute the temperature dependent phase diagram of the Heisenberg model. For this purpose we use the results from our qVQT-calculation above at $\beta=1$ and monitor the energy spectrum and calculate the correlation functions of each eigenstate independently. The correlation functions $c_0$ and $c_1$ are then obtained by mixing the correlation functions of the eigenstates according to the probabilities given in equation \eqref{eq:definition_Gibbs_state}. This approach has the advantage that the temperature-dependent results can be estimated without explicitly performing  an optimization at each temperature, but comes with the drawback that the precision decreases when the temperature differs significantly from the original optimization temperature.

This behavior is reflected in Fig.~\ref{fig:phasediag_T_theoexp}, which shows the phase boundaries as a function of temperature computed from our numerical experiments and the exact results obtained by diagonalization of the Hamiltonian. Note that the agreement is excellent for $\beta\approx1$, while the accuracy decreases as we deviate from this inverse temperature. Nevertheless, the overall phase behavior is captured correctly, demonstrating that our extension from the qVQT-calculation at $\beta=1$ captures the physics accurately.

\section{Conclusions}

In summary, we present a new variational quantum algorithm, called qVQT, which is an extension of the VQE to finite temperatures. Our approach expands on the idea of the hVQT, but implements both the entropic and energetic contribution to the free energy on a quantum circuit. In this way we effectively reduce communication between classical and quantum device and the number of executed quantum circuits to compute an accurate Gibbs state.

We demonstrate the utility of the qVQT by performing extensive numerical experiments on quantum simulators for two model systems and show that our algorithm is well suited to calculate finite temperature properties or excited states on a quantum computer. The resource requirements as well as the scaling behavior are comparable to VQEs (see also appendix~\ref{sec:apx_scaling_FVQT}), and we expect our algorithm to perform equally well for a given problem size. Hence, the qVQT provides a powerful tool to study quantum systems at finite temperature, producing useful results with resources available on current NISQ devices for a wide range of applications.

\section{Acknowledgements}
We thank the members of the MANIQU consortium for valuable expert discussions. JS, MA and TE gratefully acknowledge support from the German Federal Ministry of Education and Research (BMBF) under project No. 13N15574.

%\bibliography{literature}% Produces the bibliography via BibTeX.
%
% ****** End of file apssamp.tex ******

\clearpage

\appendix

\section{Derivation of the Measurement Precision}\label{sec:DervMeasPrec}
To compute the error of the thermal energy
\begin{align}
E = \frac{1}{Z}\sum_{i} e^{-\beta \epsilon_i}\epsilon_i = \sum_{i} p_i \epsilon_i
\end{align} 
from $N$ measurements, which are equally distributed among the eigenstates, we use the approximation of error propagation, which states that the standard deviation of a function is given by the partial derivatives and standard deviations of its arguments:

\begin{equation}
\sigma_E = \sqrt{\sum_{i}\left(\frac{\partial E}{\partial \epsilon_i}\sigma_i\right)^2}.
\end{equation}

This relation is exact in the case where all $\epsilon_i$ are independent variables. The statistical errors produced by the measurement on the quantum computer are independent and hence justifies this error estimation.
The partial derivative of the energy is:

\begin{equation}
\frac{\partial E}{\partial \epsilon_i} = p_i(1-\beta \epsilon_i)+p_i^2\beta \epsilon_i= p_i\left(1-\beta \epsilon_i(1-p_i)\right),
\end{equation} 
resulting in the standard error of the thermal energy:
\begin{align}
\Delta E = \sqrt{\frac{2^n}{N}\sum_{i} \sigma_i^2p_i^2\left(1-\beta \epsilon_i(1-p_i)\right)^2}.
\end{align}
The factor $2^n/N$ stems from the fact that we measure the eigenenergy $\epsilon_i$ with $N/2^n$ shots.

With the qVQT the probabilities are obtained from the intermediate measurement. Its standard error is obtained from summing over all possible outcomes of $N$ measurements:
\begin{equation}
\sigma_{p_i} = \sum_{j=0}^{N} \binom{N}{j}p_i^j(1-p_i)^{(N-j)}\cdot\left(\frac{j}{N}-p_i\right)^2 = \frac{p_i(1-p_i)}{N},
\end{equation}
resulting in the standard error of $p_i$ and $\epsilon_i$:
\begin{equation}
\begin{aligned}
\Delta p_i &= \frac{\sqrt{p_i(1-p_i)}}{N}\\
\Delta \epsilon_i &= \frac{\sigma_i}{\sqrt{p_iN}}.
\end{aligned}
\end{equation}

The standard error of $p_i\epsilon_i$ is calculated using the squared relative errors:
\begin{equation}
\begin{aligned}
\Delta [p_i\epsilon_i]^2 &= p_i^2\epsilon_i^2\left(\frac{\sigma_i^2}{\epsilon_i^2p_iN}+\frac{p_i(1-p_i)}{p_i^2N^2}\right)\\
&=\frac{1}{N}\left(\sigma_i^2p_i+\frac{p_i(1-p_i)}{N}\epsilon_i^2\right).
\end{aligned}
\end{equation}

The sum of these errors yields the error of the thermal energy:
\begin{align}
\Delta E = \frac{1}{\sqrt{N}}\sqrt{\sum_{i} \left[\sigma_i^2p_i+\frac{p_i(1-p_i)}{N}\epsilon_i^2\right]}.
\end{align}

\section{Number of Iterations for the 1D Heisenberg Chain}\label{sec:apx_scaling_chain}

\begin{figure}[h!]
	\includegraphics[scale=0.6]{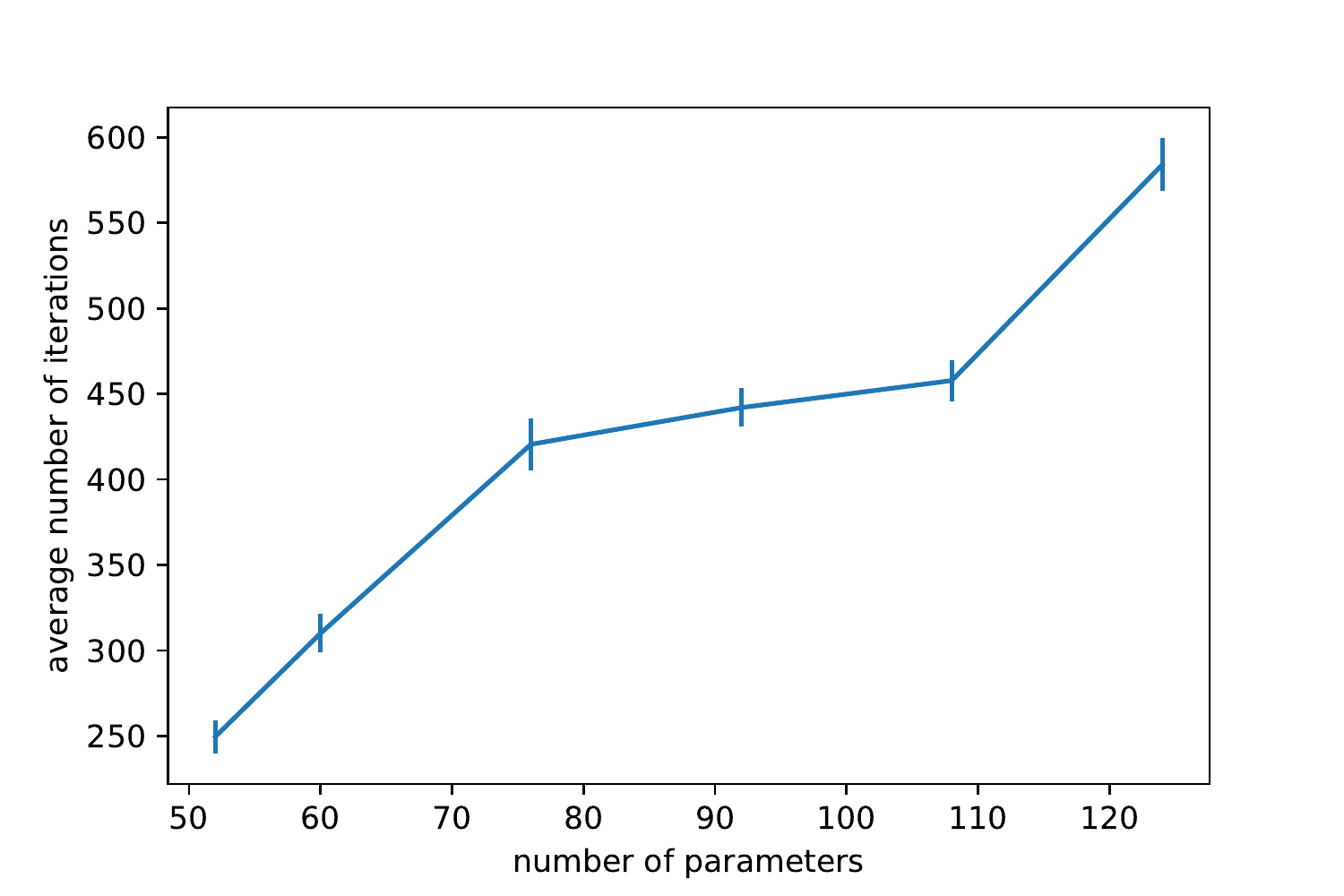}% Here is how to import EPS art
	\caption{\label{fig:4q_niter_numvar} Results of a 4-qubit 1D Heisenberg chain with transverse fields, showing the average number of iterations as a function of the total number of variational parameters.  The results were obtained from statistical averages over 100 runs starting with random initial parameters, using the L\_BFGS\_B optimizer until the gradients converged to a value of less than \num{1e-3}.}
\end{figure}

In Fig.~\ref{fig:4q_niter_numvar} we show the required number of iterations to achieve a gradient smaller than \num{1e-3} for the one-dimensional Heisenberg chain with four qubits. One iteration includes both the evaluation of the free energy together with its gradient. Using, e.g., the parameter-shift rule to obtain the gradient~\cite{schuld_evaluating_2019}, a single measurement requires the evaluation of $2n_\text{var}+1$ quantum circuits, where $n_\text{var}$ is the number of variational parameters. The data presented in Fig.~\ref{fig:4q_niter_numvar} suggests that the number of iterations is roughly linear in the number of parameters, i.e., the cost of a qVQT scales quadratically with the number of parameters. Note that our results are obtained using the L\_BFGS\_B optimizer of qiskit and could in general depend on the choice of the optimizer. However, the number of iterations scales in the same manner as long as the optimization algorithm shows the same convergence behavior as the L\_BFGS\_B algorithm with respect to the condition number.  

\section{Scaling of the qVQT with System Size}\label{sec:apx_scaling_FVQT}
\begin{figure}
	\includegraphics[scale=0.6]{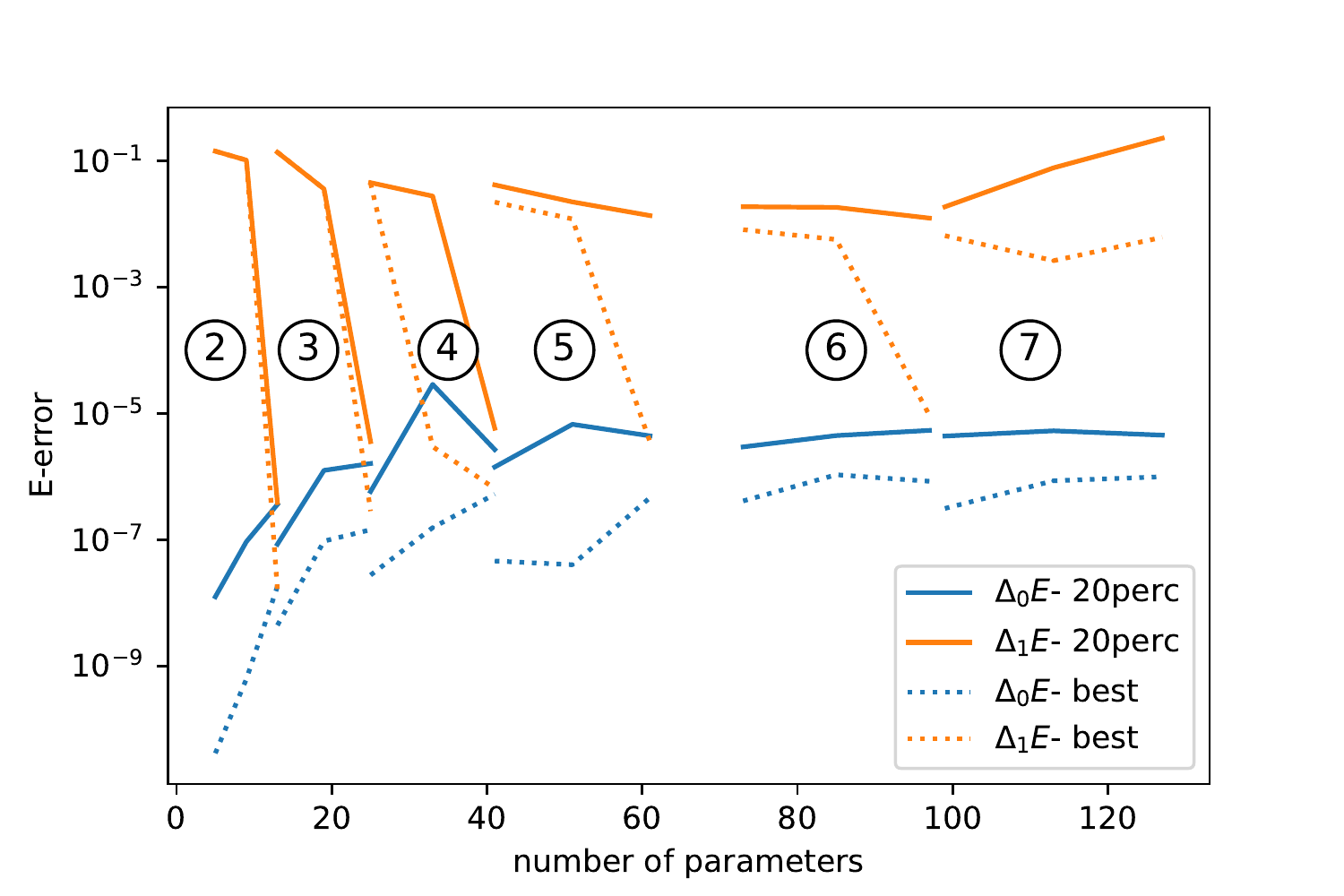}% Here is how to import EPS art
	\caption{\label{fig:HB_chain_scaling} Results for a 1D Heisenberg chain with transverse fields, showing the two first energy differences depending on the total number of parameters for different lengths of the chains $N$ (indicated by the encircled integer numbers, \textcircled{\raisebox{-0.9pt}{$N$}}). The results were obtained from statistical averages over 100 runs starting with random initial parameters, using the L\_BFGS\_B optimizer until the gradients converged to a value of less than \num{1e-3}. The solid lines denote the \nth{20} percentile of the runs and the dotted lines show the best runs.}
\end{figure}

Here we briefly discuss the scaling of the qVQT with respect to system size by studying the 1D Heisenberg chain from section \ref{sec:1dHB} and by varying the chain length $N$. In this numerical experiment we only keep a minimal first variational circuit $\text{VQC}_1$, which only contains one rotation on the first qubit, and try to reproduce the ground-state and the first excited state. In Fig.~\ref{fig:HB_chain_scaling} we show the first two energy differences defined as
\begin{equation}
	\Delta_kE = \frac{\sum_{i=0}^{k}p_i|\epsilon_i^\text{exp}-\epsilon_i^\text{exact}|}{n\sum_{i=0}^{k}p_i},
\end{equation}
which are the weighted energy differences between the experimental eigenstates and the corresponding exact values, divided by the number of qubits $n$.

Based on our results, the required circuit depth for the second variational circuit seems to grow linearly with the chain length. The number of parameters per circuit layer additionally grows linearly with the number of qubits, which yields an overall quadratic scaling of the total number of parameters. For the results with 5, 6 and 7 qubits we see that the convergence of the classical optimization limits the quality of the results, but the variational circuit is able to reproduce the correct states. This behavior is comparable to the VQE where for growing number of qubits the classical optimization becomes increasingly challenging~\cite{bittel_training_2021}.

\end{document}